\newif\ifpreprint

\preprintfalse 

\ifpreprint
\documentclass[journal=jctccd,manuscript=article]{achemso}
\setkeys{acs}{articletitle=true}
\else
\documentclass[journal=jctccd,manuscript=article,layout=twocolumn]{achemso}
\setkeys{acs}{articletitle=true}
\fi
\usepackage[usenames,dvipsnames]{xcolor}                        
 
\definecolor{JCTCgreen}{RGB}{29,111,66}
\definecolor{JCTCblue}{RGB}{40,68,130}
\usepackage[colorlinks = true,
            linkcolor = blue,
            urlcolor  = black,
            citecolor = JCTCblue,
            anchorcolor = black]{hyperref}
                                                                                                                                                   
\usepackage{amsmath}                                                                                                                               
\usepackage{amssymb}                                                                                                                               
\usepackage{amsfonts,amsthm}
\usepackage[mathscr]{euscript}                                                                                                                     
\usepackage{graphicx}                                                                                                                              
\usepackage{dcolumn}                                                                                                                               
\usepackage{bm}                                                                                                                                    
\usepackage{braket}                                                                                                                                
\usepackage{multicol}                                                                                                                              
\usepackage{dcolumn}                                                                                                                               
\usepackage{multirow}                                                                                                                              
\usepackage{subcaption}                                                                                                                            
\usepackage{ifthen}                                                                                                                                
\usepackage{booktabs}                                                                                                                              
\usepackage{xspace}                                                                                                                                
\usepackage{color,soul}                                                                                                                          
                                                                                                                                                   
\usepackage{threeparttable}                                                                                                                        
                                                                                                                                                   
\usepackage{footnote}                                                                                                                              
\makesavenoteenv{tabular}                                                                                                                          
\makesavenoteenv{table}                                                                                                                            
                                                                                                                                                   
\usepackage[version=3]{mhchem} 
\usepackage[T1]{fontenc}       
\usepackage{lineno}                                                                                                                                
\modulolinenumbers[2]                                                                                                                              
\usepackage{titlesec}                                                                                                                              
\usepackage{footnote}

\ifpreprint
\usepackage[fontsize=12pt]{scrextend} 
\else
\usepackage[fontsize=10pt]{scrextend}
\fi
                                                                                                             
\titleformat{\section}
  {\normalfont\fontfamily{phv}\selectfont\bfseries\color{JCTCgreen}}
  {\thesection}{0.25em}{\uppercase}
  
  \titleformat{\subsection}[runin]
  {\normalfont\fontfamily{phv}\selectfont\bfseries\color{black}}
  {\quad\thesubsection}
  {0.75em}
  {}[.\;\;]
  
    \titleformat{\subsubsection}[runin]
  {\normalfont\fontfamily{phv}\selectfont\itshape\color{black}}
  {\quad\thesubsubsection}
  {0.75em}
  {}[.\;\;]

  \titleformat{\suppinfo}
  {\normalfont\sffamily\bfseries}
  {\thesubsection}{0.25em}{.}
    
\renewcommand{\thesection}{\arabic{section}.}
\renewcommand{\thesubsection}{\arabic{section}.\arabic{subsection}.}
\renewcommand{\thesubsubsection}{\arabic{section}.\arabic{subsection}.\arabic{subsubsection}.}                                                                                                                                                                                                 

\newcommand*{\kcal}{kcal~mol$^{-1}$\xspace}
\newcommand*{\Eh}{$E_\mathrm{h}$\xspace}
\newcommand*{\sunit}{$E_\mathrm{h}^{-2}$\xspace}
\newcommand{\cre}[1]{\hat{a}_{#1}^{\dagger}}
\newcommand{\ann}[1]{\hat{a}_{#1}}

\usepackage{ifthen}

\definecolor{goodorange}{RGB}{225,125,0}
\definecolor{goodgreen}{RGB}{0,125,0}
\definecolor{goodred}{RGB}{220,50,25}
\definecolor{goodblue}{RGB}{25,25,150}

\newcommand{\note}[2]{
\ifthenelse{\equal{#1}{F}}{
\colorbox{goodorange}{\textcolor{white}{\footnotesize \fontfamily{phv}\selectfont #1}}
    \textcolor{goodorange}{{\footnotesize \fontfamily{phv}\selectfont #2}}
}{}
\ifthenelse{\equal{#1}{J}}{
\colorbox{goodgreen}{\textcolor{white}{\footnotesize \fontfamily{phv}\selectfont #1}}
    \textcolor{goodgreen}{{\footnotesize \fontfamily{phv}\selectfont #2}}
}}

\let\oldmaketitle\maketitle                                                                                                                                                          
\let\maketitle\relax 

\title{A Combined Selected Configuration Interaction and Many-Body Treatment of Static and Dynamical Correlation in Oligoacenes}

\author{Jeffrey B. Schriber}
\affiliation{Department of Chemistry and Cherry L. Emerson Center for Scientific Computation, Emory University\\Atlanta, Georgia, 30322, USA}

\author{Kevin P. Hannon}
\affiliation{Department of Chemistry and Cherry L. Emerson Center for Scientific Computation, Emory University\\Atlanta, Georgia, 30322, USA}

\author{Francesco A. Evangelista}
\affiliation{Department of Chemistry and Cherry L. Emerson Center for Scientific Computation, Emory University\\Atlanta, Georgia, 30322, USA}
\email{francesco.evangelista@emory.edu}

\begin{document}

\ifpreprint
\linenumbers                                                                                                                                                                          
\else                                                                                                                                                         
\twocolumn[
\begin{@twocolumnfalse}                                                                                                                                                             
\fi  

\oldmaketitle

\begin{abstract}
We have combined our adaptive configuration interaction (ACI) {[J.B. Schriber and F.A. Evangelista, J. Chem. Phys. \textbf{144}, 161106 (2016)]} with 
a density-fitted implementation of the second-order perturbative multireference driven similarity renormalization group (DSRG-MRPT2) {[K.P. Hannon, C. Li and F.A. Evangelista J. Chem. Phys. \textbf{144}, 204111 (2016)]}.
We use ACI reference wave functions to recover static correlation for active spaces larger than the conventional limit of 18 orbitals. The dynamical 
correlation is computed using the DSRG-MRPT2 to yield a complete treatment of electron correlation. 
We apply the resulting method, ACI-DSRG-MRPT2, to predict singlet-triplet gaps, metrics of open-shell character, and spin-spin correlation functions  for the oligoacene series (2--7 rings).
Our computations employ active spaces with as many as 30 electrons in 30 orbitals and up to 1350 basis functions, yielding gaps that are in good agreement with available experimental results.
Large bases and reference relaxation lead to a significant reduction in the estimated radical character of the oligoacenes with respect to previous valence-only treatments of correlation effects.
\end{abstract}

\ifpreprint 
\else      
\end{@twocolumnfalse} 
]   
\fi 

\section{Introduction}

In order to make meaningful predictions about the electronic properties of molecules,
theoretical computations need to satisfactorily converge both correlation and basis set effects.
By correlation, we refer to its normal decomposition into a \emph{static} component, defined by the strong mixing of electronic configurations typically resultant from degeneracies, and
a \emph{dynamical} component, which 
includes short-range Coulombic and long-range dispersion interactions.\cite{DanielKWMok:1996ki,Evangelista:2018bt}
Molecules with more than 2-4 strongly correlated electrons require a multireference method,
wherein a specialized approach is adopted for both types of correlation. 
Static correlation is treated rigorously with a multiconfigurational reference wave function defined in a set of active orbitals, 
commonly from complete active space configuration interaction (CASCI) or CAS self consistent field (CASSCF),\cite{Roos:1980ba,Werner:1998dh,Olsen:2011kv} 
while more affordable perturbative or non-perturbative many-body theories are invoked for dynamical correlation.
Unfortunately, both correlation treatments will fail once the number of active orbitals becomes too large, usually around 18 active orbitals for CASCI,
relegating applications of multireference methods to regimes in chemistry where static correlation is defined in just a few orbitals.
In this article, we introduce a new strategy capable of treating static and dynamical correlation using large active spaces 
and basis sets to approach chemical accuracy ($\leq 1$ \kcal error).

The limitations of CASSCF and CASCI arise from 
the number of variational parameters growing combinatorially as the number of electrons and orbitals in the active space. 
New techniques can achieve a sub-combinatorial cost, or a significant reduction of the prefactor, by exploiting the sparse 
structure of CASCI wave functions, enabling reliable computations using significantly larger active spaces. \cite{white1992density,white1999ab,chan2002highly,chan2008introduction,chan2011density,DePrinceIII:2016kx,FossoTande:2016cn,FossoTande:2016hb,booth2009fermion,booth2010approaching,cleland2010communications,li2016combining,Sharma:2017iu,holmes2016heat,Smith:2017fm,tubman2016deterministic,schriber2016communication,Schriber:2017jd,Ohtsuka:2017cw,Zhang:2016gu,Vogiatzis:2015km,Fales:2018gb,Garniron:2018hl,
Tubman:vd,Guo:2018ct,Pathak:2017cz,Xu:wz}
However, when augmenting any CAS method with a theory of dynamical correlation, numerical and practical concerns arise.
Most multireference theories suffer from numerical instabilities known as ``intruder states'' which
spoil computations with unphysical contributions difficult to systematically cure.\cite{Roos:1995jz}  
Additionally, theories of dynamical correlation require the computation of high-order density matrices and cumulants, 
causing methods like CAS second-order perturbation theory (CASPT2)\cite{Andersson:1998cq} to scale as the ninth power
of the number of active orbitals, limiting these methods to moderate active space sizes (<20-25). 
The development of approximate CAS techniques has as of yet outpaced the ability for dynamical correlation
methods to make use of them.

Some recent work has connected large active space ($\geq 20$ orbitals) wave functions with dynamical correlation treatments,
including active space DFT approaches\cite{Manni:2014kg,Ghosh:2017gl,Gagliardi:2017bu,Fromager:2007id,Sharmadmrgdft}
 and the combination of density matrix renormalization group with
perturbative\cite{Zgid:2009fu,kurashige2014complete,Kurashige:2014uf} and CI-based\cite{Luo:2018gn} theories.
Recently, the adiabatic connection has been used to compute correlation energies from CAS wave functions and
only requires two-body density matrices.\cite{Pastorczak:2018fl,Pernal:2018fg}
To our knowledge, none of these approaches have treated the coupling of static and dynamical correlation, which
we refer to as the \emph{relaxation} effect of the reference wave function in response to a dynamical correlation treatment, 
the importance of which is not fully understood.
For this reason, and other drawbacks related to cost, density functional dependence, and
potential neglect of static correlations, the development of a systematically improvable,
numerically stable, and computationally efficient multireference theory with dynamical correlation is still an open problem. 

\section{Theory}
We seek a complete treatment of electron correlation by, for the first time, interfacing two of our recently developed theories, 
the adaptive CI (ACI)\cite{schriber2016communication,Schriber:2017jd} and the second-order perturbative variant of the 
multireference driven similarity renormalization group (DSRG-MRPT2).\cite{Evangelista:2014kt,Li:2015iz,Li:2016hb,Hannon:2016bh,Li:2017ff}

\subsection{Adaptive CI}
The goal of the ACI procedure is to converge an approximate CASCI wave function,
\begin{equation}
\ket{\Psi_M} = \sum_{\Phi_{\mu}\in M} C_{\mu}\ket{\Phi_{\mu}}
\end{equation}
in an iteratively selected determinantal model space, $M$, such that the variational energy error approximately matches a user defined 
parameter, $\sigma$, i.e.
\begin{equation}
|E_{\rm{CASCI}} - E_{M}| \approx \sigma \label{eq:cond}
\end{equation}
where $E_{M} = \braket{\Psi_M | \hat{H} | \Psi_M}$ is the ACI energy.
The ACI wave function is built by iteratively growing a set of reference determinants ($P$) and screening its first order interacting space ($F$)
using perturbative energy estimates. To ensure error control, the screening is done by excluding determinants with the smallest
energy estimates such that the accumulation of these perturbative corrections approximately equals the energy criterion, $\sigma$.
Additional details of the algorithm are presented elsewhere,\cite{schriber2016communication,Schriber:2017jd}
but we emphasize that the iterative procedure rigorously samples the total
CASCI wave function to very closely meet the condition of eq \eqref{eq:cond} and yield 
a compact wave function whose energy error is controlled \textit{a priori} by the user.
As a result, ACI wave functions for different electronic states 
can be made with near-equal accuracy to approach perfect error cancellation 
in relative properties, such as singlet-triplet splittings.\cite{schriber2016communication,Schriber:2017jd}
We have previously illustrated this error cancellation in systems using up to 42 active orbitals.\cite{schriber2016communication}

\begin{figure*}[ht!]
\centering
\includegraphics[width=6in]{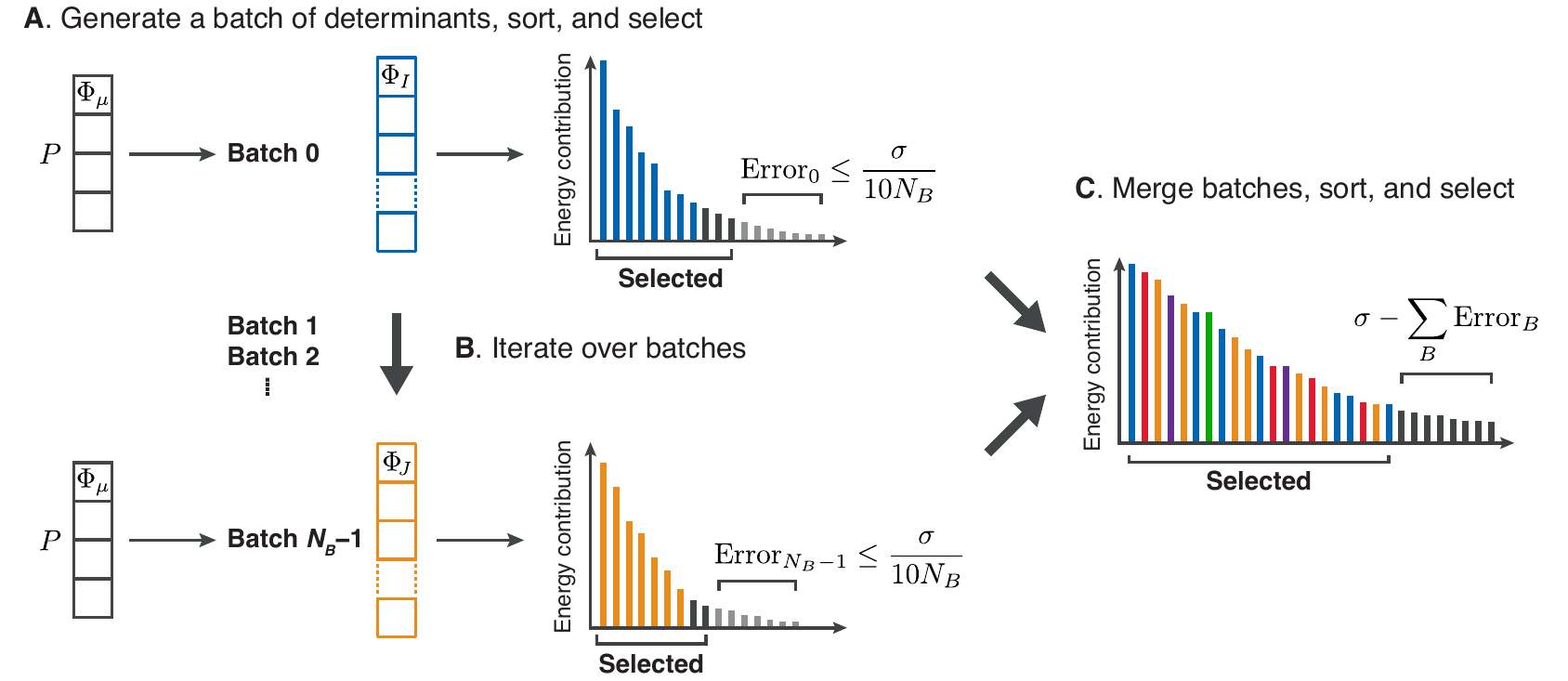}
\caption{Summary of batched screening algorithm. (\textbf{A}) Determinants in the first batch are generated from a subset of the single and double excitations from the reference $P$ space. The excited determinants in the batch are sorted and screened using a scaled $\sigma$ value, with only selected ones being stored. (\textbf{B}) The procedure in \textbf{A} is repeated for all batches sequentially, with all selected determinants stored. (\textbf{C}) These selected determinants are merged and sorted, and a final screening is done to ensure that the total correlation energy ignored corresponds to the original $\sigma$ value.}
\label{fig:screen}
\end{figure*}

In our previous implementation, all determinants $\Phi_{I}$ in $F$ were stored simultaneously
during the screening step of the algorithm, which incurs a memory cost of
$\mathcal{O}(|P|N_{O}^{2}N_{V}^{2})$ where $|P|$ is the number of reference determinants in the set $P$, 
and $N_{O}$ and $N_{V}$ are number of the occupied and virtual orbitals in the active space, respectively. 
In addition to being the storage bottleneck, screening is usually the most expensive step with a complexity of $\mathcal{O}(|P|N_{O}^{2}N_{V}^{2}\log(|P|N_{O}^{2}N_{V}^{2}))$ 
due to sorting of all energy estimates.
This storage becomes prohibitive if low values of $\sigma$ are used in large active space computations. 
We have now implemented a batched selection algorithm that can mitigate 
this memory bottleneck by sequentially screening subsets of $F$ using a scaled $\sigma$ value, summarized in Figure \ref{fig:screen}. 
The number of batches, $N_{B}$, is chosen to be the smallest value such that an approximate estimate of the memory requirement of $\frac{|F|}{N_{B}}$
can be handled by a single node.
For each batch $B \in \{0, 1, \ldots, N_{B} - 1\}$, we iterate through all determinants in $P$, compute a hash value for each single and double excitation, $h(\Phi_{I})$,
and store this determinant with its associated energy estimate only if $h(\Phi_{I})\mod B = 0$.
With the resulting subset of $F$, we do our conventional screening described previously, but scaling $\sigma$ by $\frac{1}{10N_{B}}$
to avoid over-truncation of each batch, where the factor of $\frac{1}{10}$ is determined empirically.
Once all batches have been screened, the surviving determinants are merged, and the
accumulated energy estimates of determinants excluded are summed among batches to obtain an estimate of the total correlation being ignored. 
With the merged determinants, we do a final screening with $\sigma$ shifted by the energy already screened from the batches. The remaining
determinants merged with the reference $P$ define the full model space $M$.
Importantly, our hash function is evenly distributed among determinants
so each batch contains fairly equally-weighted determinants and is thus able to exactly match results of the original algorithm.
What results is memory storage of $\mathcal{O}(\frac{1}{N_{B}}|P|N_{O}^{2}N_{V}^{2})$ and a reduced complexity of $\mathcal{O}(|P|N_{O}^{2}N_{V}^{2}\log(\frac{1}{N_{B}}|P|N_{O}^{2}N_{V}^{2}))$.

\subsection{Implementation of RDMs}
To connect ACI reference wave functions to the DSRG-MRPT2 perturbative treatment,
the 1-, 2-, and 3- body reduced density matrices (RDMs) are required. Introducing second quantized creation ($\hat{a}^\dagger$) and annihilation ($\hat{a}$) operators, a generic $k$-particle RDM ($\gamma_{rs\dots}^{pq\dots}$) may be expressed as
\begin{equation}
\gamma_{rs\dots}^{pq\dots}  = \sum_{\Phi_{I},\Phi_{J} \in M} \bra{\Phi_{I}}\cre{p}\cre{q}\dots\ann{s}\ann{r}\ket{\Phi_{J}}C_{I}C_{J} \label{eq:rdm}
\end{equation}
Storage and contraction of the 3-RDM scales only as $\mathcal{O}(N_{A}^{6})$ for the 3-RDM, where $N_{A}$ is the number of active orbitals.
This moderate scaling enables the use of large active spaces unreachable with conventional many-body dynamical correlation 
methods, for which active spaces beyond 20-24 active orbitals are impossible to treat without invoking 
approximations that can potentially introduce artificial numerical problems.\cite{Zgid:2009fu}

Despite this reduced scaling, the construction of the 3-RDM from a 
selected CI wave function is itself a formidable task. As is commonly done in selected CI diagonalization
procedures, intermediate residue lists, which map all Slater determinants in a set to all possible
determinants with one or two fewer electrons,\cite{Stampfuss:2005bk} can be used to predetermine all non-zero elements of the Hamiltonian matrix
or of the RDMs. Our ACI implementation adopts this strategy for the 1- and 2- RDMs and in directly building the sigma vector $\boldsymbol{\sigma} = \mathbf{H}\mathbf{c}$
during diagonalization, as the storage scales at most as
$\mathcal{O}(N_\mathrm{det}N_\mathrm{el}^{2})$ for $N_\mathrm{det}$ determinants and $N_\mathrm{el}$ active electrons.
For constructing the 3-RDM, storage of these lists become impractical 
since triple annihilations are required, increasing the memory scaling to $\mathcal{O}(N_\mathrm{det}N_\mathrm{el}^{3})$.
These lists also become prohibitive for the lower RDMs and for building $\boldsymbol{\sigma}$ for very large ($\geq 10^{7}$) determinant spaces. 
We have implemented a more memory efficient algorithm similar to ref \citenum{Sharma:2017iu},
where we organize determinants by common occupation strings of $\alpha$ or $\beta$ electrons. 
For components of the RDMs whose indices all correspond to the same spin, an element is easily computed by looping
over determinants only with the same occupation in the opposite spin string, and performing a bitwise comparison to determine
the appropriate creation/annihilation indices for evaluating equation \eqref{eq:rdm}.
For mixed spin components, a loop over all $\alpha$ strings is required to determine which strings differ by the desired number of occupation differences. Then, a double loop only over determinants containing those $\alpha$ strings with the correct number of substitutions for the particular RDM component is performed to 
compute eq \eqref{eq:rdm}. Thus, the storage requirement changes only to twice the number of determinants, generally less
than the 3-RDM itself, despite
formally costing $\mathcal{O}(N^2_O N^2_V  N_{\alpha}K^2_{\alpha} + N^2_O N^2_V  N_{\beta} K^2_{\beta})$ in computational time, with $N_{\alpha}/N_{\beta}$ referring
to the number of $\alpha/\beta$ strings, and $K_{\alpha}/K_{\beta}$ the average number of determinants per $\alpha/\beta$ string.

\subsection{DSRG-MRPT2}
All theories based on the MR-DSRG avoid intruder states by gradually block-diagonalizing 
the Hamiltonian ($\hat{H}$) using a unitary transformation dependent on a continuous flow parameter, $s$,
\begin{equation}
\hat{H}\rightarrow\bar{H}(s) = \hat{U}(s)\hat{H}\hat{U}^{\dagger}(s) \quad s \in [0,\infty) \label{eq:flow}
\end{equation}
The unitary operator is written in a connected form as $\hat{U}(s)= e^{\hat{A}(s)} = e^{\hat{T}(s) - \hat{T}^{\dagger}(s)}$, where the cluster operator $\hat{T}(s)$ is analogous to the coupled cluster operator.\cite{Evangelista:2011eh,Hanauer:2012hn}
In the DSRG, $\hat{T}(s)$ is determined implicitly by solving a set of nonlinear equations\cite{Evangelista:2014kt}
\begin{equation}
 [\bar{H}(s)]_{\rm{N}}=\hat{R}(s)  \label{eq:source}
\end{equation}
where the source operator, $[\hat{R}(s)]$, drives the transformation, and the many-body condition expressed by eq ~\eqref{eq:source} implies exclusive inclusion of nondiagonal (N) terms which couple the reference to its excited configurations.
The DSRG total energy is then computed as, 
\begin{equation}
E(s) = \bra{\Psi_{M}}\bar{H}(s)\ket{\Psi_{M}} \label{eq:energy}
\end{equation}
Performing an order-by-order expansion of eqs \eqref{eq:flow}--\eqref{eq:energy}, 
we can write the second-order MR-DSRG energy\cite{Li:2015iz} as,
\begin{equation}
E^{(2)}(s) = \frac{1}{2}\bra{\Psi_{M}}[\tilde{H}^{(1)}(s),\hat{A}^{(1)}(s)]\ket{\Psi_{M}} \label{eq:ept2}
\end{equation}
where the modified first-order Hamiltonian, $\tilde{H}^{(1)}(s)$, is determined by the source operator, 
$\tilde{H}^{(1)}(s) = \hat{H}^{(1)}(s) + [\hat{R}^{(1)}(s)]_{\rm{N}}$.

We will report the DSRG-MRPT2 energy from two procedures.
Firstly, we can compute the expectation
value defined in eq \eqref{eq:energy} by coupling the reference, defined in a set of active
orbitals, with the full set of non-frozen occupied and virtual orbitals using eq \eqref{eq:ept2}. 
This energy is the \emph{unrelaxed} energy since the reference wave function is constant, and the corresponding approach can be classified as a diagonalize-then-perturb method.
Alternatively, the first-order effective Hamiltonian, $\bar{H}^{(1)}(s)$, can be rediagonalized to yield 
the \emph{relaxed} energy and wave function 
in a diagonalize-perturb-diagonalize scheme, each diagonalization done with ACI.

The combination of ACI and DSRG represents one of the few viable options for accurately studying large molecules with complex electronic structures. 
ACI allows us to use large active spaces while the DSRG provides an efficient, systematically
improvable, and intruder-free formalism to recover dynamical correlation. 
Beyond energies, 
the relaxation procedure computes reference wave functions in response to dynamical correlation effects,
a feature neglected with CASPT2 and DFT approaches.
These relaxed wave functions allow us to probe electronic properties without inherent bias towards effects of static correlation.

\section{The Oligoacenes}
This balanced treatment of static and dynamical correlation is particularly important in our application to the oligoacenes, 
or $n$-acenes for $n$ linearly fused benzene rings. The oligoacenes have long been of 
fundamental interest to chemists due to their semiconducting and optical properties,\cite{Bettinger:2010go,Yang15082016,Paci:2006kn,Marciniak:2007ba,Kuhlman:2010gn,Zimmerman:2011ch,Wilson:2013gr,Izadnia:2017dn,Wibowo:2017jr,RefaelyAbramson:2017gw,MartinezMartinez:2018ea}
and theoreticians have been studying their singlet-triplet splittings
and ground state electronic structures with particular attention to the 
disputed emergence
of a stable, open-shell singlet ground state with increasing acene size.
\cite{Bettinger:2010go,Bendikov:2004fg,Hachmann:2007ft,Hajgato:2009dw,Hajgato:2011jg,Mizukami:2013ed,Malrieu:2014fy,Lehtola:2017il,Lehtola:2016bb,Yeh:2016kn,FossoTande:2016cn,
schriber2016communication,Lee:2017bh,Battaglia:2017bb,Dupuy:2018gp}
In characterizing the ground state, qualitatively different interpretations can arise depending on the
degree dynamical correlation included,
with recent studies suggesting that pure active-space methods tend to overestimate
biradical character in these ground states.\cite{Lehtola:2017hf,Lee:2017bh,Battaglia:2017bb,Dupuy:2018gp} 
An accurate theoretical characterization of the oligoacenes is complicated, however,  by 
($i$) strong correlation in the $\pi/\pi^{*}$ manifold, 
($ii$) their size, prohibitive for many \textit{ab initio} methods, and 
($iii$) the large basis sets required for experimental comparisons.
For these reasons, a chemically accurate prediction of the singlet-triplet splittings 
and precise descriptions of the ground states have remained elusive to theoretical techniques. 
The application of ACI-DSRG represents an important step in understanding fundamental electronic properties of oligoacenes.

\subsection{Computational Details}
Singlet and triplet state geometries for the oligoacenes, reported in the supplemental information, were optimized at the UB3LYP/6-31G($d$) level of theory and are generally the same as those reported by Hachmann and co-workers.\cite{Hachmann:2007ft}.
Analytic Hessians were computed with ORCA\cite{neese2012orca}
to ensure stability of our geometries and to compute zero-point vibrational energy corrections (ZPVEs). 
The ACI-DSRG-MRPT2 approach was implemented in \caps{Forte}, our freely available software, which is run as a plugin to \caps{Psi4}.\cite{Parrish:2017hga}
All ACI-DSRG-MRPT2 computations use restricted Hartree-Fock (RHF) and restricted open-shell HF (ROHF) orbitals for singlet and triplet computations, respectively,
and the full
$\pi/\pi^{*}$ manifold in the active space, resulting in a CAS($4n+2,4n+2$), for $n$ fused benzene rings.
We use the conventional notation of CAS($e$,$o$), for $e$ active electrons and $o$ active orbitals.
To further compress CI expansions, we separately localize doubly occupied, singly occupied, and virtual
active orbitals,\cite{Bytautas:2003un} 
which requires the use of $C_{1}$ symmetry.
The ACI computations of the reference use a prescreening threshold of $\tau_{V} = 10^{-12}$ \Eh  (see ref \citenum{Schriber:2017jd}) for all acenes
except hexacene which required $\tau_{V} = 10^{-7}$ \Eh, which is still safely above the corresponding value of $\sigma$.
Our ACI-DSRG-MRPT2 computations are run in the cc-pV$X$Z, ($X = $ D, T, Q) basis sets,\cite{DunningJr:1989bx,Woon:1993in}
 with all $1s$-like
orbitals on carbon atoms treated with the frozen core approximation. For all computations, we use density fitted integrals
and a DSRG-MRPT2 implementation specialized for three-index integrals.\cite{Hannon:2016bh}
We use the corresponding cc-pV$X$Z-JKFIT auxiliary basis\cite{Weigend:2002ga}  for RHF/ROHF computations, 
and the cc-pV$X$Z-RI auxiliary basis sets\cite{Weigend:2002jp,Hattig:2005dm} for DSRG-MRPT2 computations of the correlation energy.

In running ACI-DSRG-MRPT2, two user-specified parameters need to be considered. The first is the energy importance criterion in ACI, $\sigma$,
which we choose to be as small as practically possible.
We will show that our reported energies are converged with respect to $\sigma$. 
The second parameter we need to select is the flow parameter, $s$, of the DSRG-MRPT2. 
Guided by our previous work with the DSRG,\cite{Li:2015iz,Hannon:2016bh} 
we use $s=0.5$ \sunit for all computations in order to recover sufficient correlation without
becoming vulnerable to intruders. 

\subsection{Singlet-Triplet Splittings}

\begin{figure*}[ht!]
\centering
\includegraphics[width=6in]{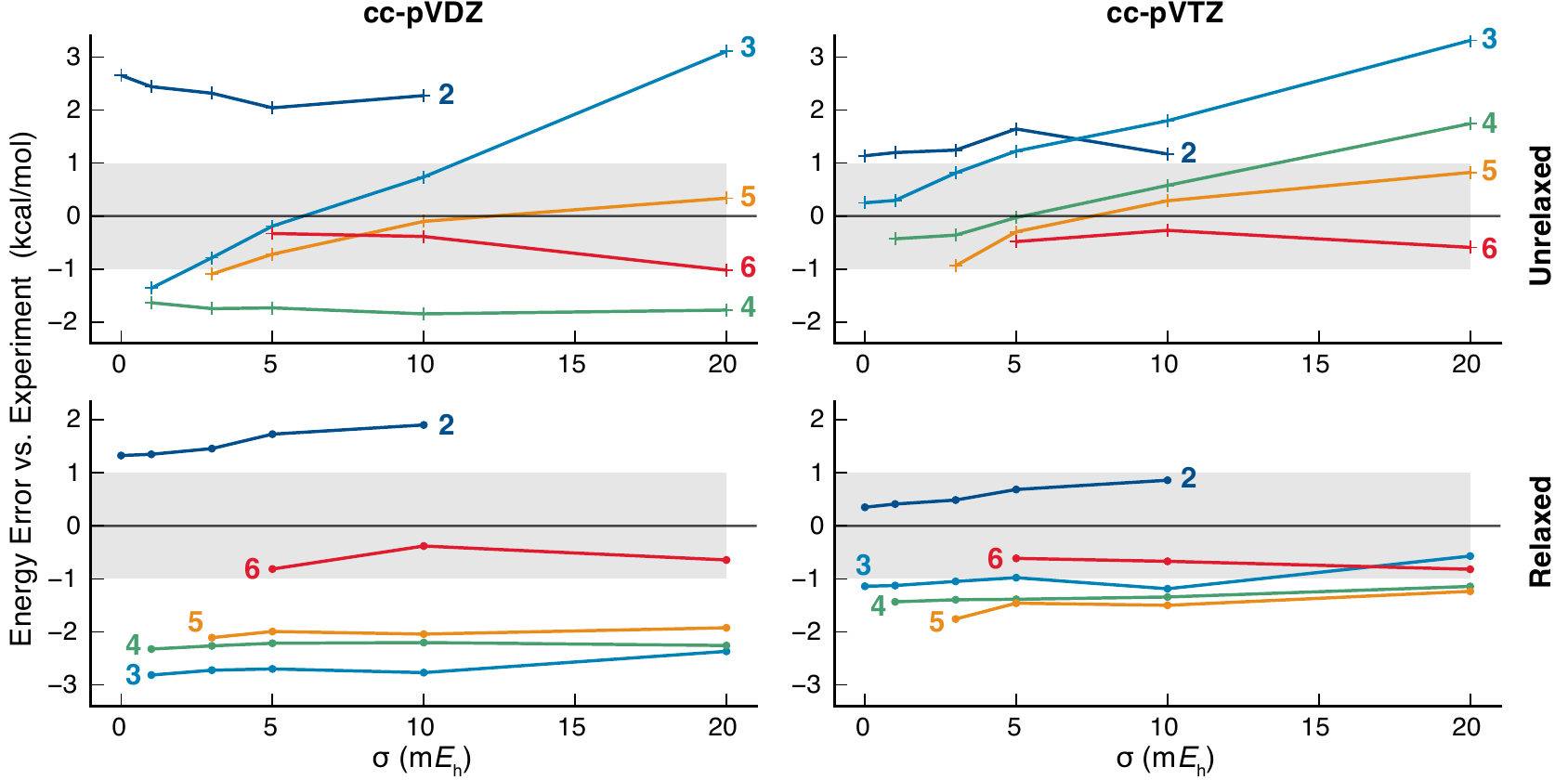}
\caption{ACI-DSRG-MRPT2 error in adiabatic singlet-triplet splitting of $n$-acenes ($2 \leq n \leq 6$) with respect to experiment using
cc-pVDZ (left) and cc-pVTZ (right) basis sets and unrelaxed (top) and relaxed (bottom) references. 
The shaded area indicates a $\pm$1 \kcal error window, and no zero-point vibrational energy correction is included.}
\label{fig:acene_error}
\end{figure*}

To understand the behavior of ACI-DSRG-MRPT2,
we first analyze the errors in the adiabatic
singlet-triplet splittings ($\Delta E_{\rm{ST}} = E_{\rm{T}} - E_{\rm{S}}$) of $n$-acenes ($n = 2-6$) with both
relaxed and unrelaxed references built with varying $\sigma$, using the cc-pVDZ and cc-pVTZ basis sets. These computations
use optimized geometries for both singlet and triplet states, and we show the
errors with respect to experimental adiabatic $\Delta E_{\rm{ST}}$\cite{Siebrand:1967dg,birks1970,Schiedt:1997hy,Burgos:1977ck,Angliker:1982jb}
in Figure \ref{fig:acene_error}.
For unrelaxed references, the convergence of the error with respect to decreasing $\sigma$ is slow and at times erratic, particularly for naphthalene computed with cc-pVDZ. Upon relaxation of the reference, we see that the errors become very 
stable, even for large $\sigma$ ($10~mE_{\rm{h}}$). This effect suggests that the reference relaxation can help 
alleviate inaccuracies from an overly-truncated reference wave function.
Even for small $\sigma$ values, relaxation effects can be large and tend to be larger 
for smaller acenes. 
Using the cc-pVTZ basis, the errors in the relaxed energies for all acenes converge with decreasing $\sigma$ and seem to loosely agree
with the experimental values despite neglect of vibrational effects and the non-exact geometries we employ. The most important
result from Figure \ref{fig:acene_error} is that, using a large basis set and a relaxation procedure, we can use highly-truncated wave functions without incurring serious energetic errors.

The singlet-triplet splittings in Figure \ref{fig:acene_error} corresponding to the smallest $\sigma$ used are 
summarized in Table \ref{tab:st} along with available cc-pVQZ data, for which the largest number of basis functions
we could treat was 1350 for tetracene in the cc-pVQZ basis.
We also report singlet-triplet splittings for heptacene, requiring a CAS(30,30), although we were restricted to $\sigma =$ 10 m\Eh
due to computational time and memory constraints. While we do see that our relaxed energies are stable with respect to an increasing $\sigma$,
the results from heptacene are possibly affected by an overly truncated reference. Moreover, the heptacene splittings are the only ones
that increase with relaxation, which can be interpreted as sign of an insufficient reference being corrected by relaxation.
Generally, we see that relaxation reduces splitting energies by up to 1 \kcal, though decreasingly so with increasing acene size.
Increasing the basis set from cc-pVTZ to cc-pVQZ increases the splittings by over 1 \kcal, indicating the importance of basis
set effects in accurately predicting these quantities.

\begin{table}[ht!]      
\begin{threeparttable}              
    \footnotesize    \caption{Adiabatic singlet-triplet splittings (\kcal) of the acene series computed with ACI and 
    ACI-DSRG-MRPT2 with both unrelaxed and relaxed references.}       
    \begin{tabular*}{\columnwidth }{@{\extracolsep{\fill}} ccccccc}   
   
    \hline  
                     
    \hline                  
  &                         &                           &     &    \multicolumn{3}{c}{$\Delta E_{\rm{ST}}$ (\kcal)} \\ \cline{5-7}  
    & & & & & \multicolumn{2}{c}{ACI-DSRG-MRPT2} \\ 
      $n$ & CAS($n$,$n$) & $N_{\rm{bf}}{}^{a}$ & $\sigma^{b}$ & ACI &  unrelaxed & relaxed \\   \hline                            
    \multicolumn{7}{c}{\bf cc-pVDZ} \\                                                                                     
    2 & (10,10)  & 170 & 0.0 & 68.3 & 63.7 & 62.3  \\      
    3 & (14,14)  & 232 & 0.0 & 43.6 & 41.7 & 40.3   \\   
    4 & (18,18)  & 294 & 1.0 & 30.9 & 27.7 & 27.0 \\                                
    5 & (22,22)  & 356 & 3.0 & 21.4 & 18.7 & 17.7 \\                                                                                            
    6 & (26,26)  & 418 & 5.0 & 13.6 & 11.7 & 11.2\\
    \multicolumn{7}{c}{\bf cc-pVTZ} \\                                                                                     
    2 & (10,10)  & 402 & 0.0 & 64.1 & 62.1 & 61.4 \\      
    3 & (14,14)  & 546 & 0.0 & 45.6 & 43.4 & 42.0\\                                                                                                  
    4 & (18,18)  & 690 & 1.0 & 33.2 & 28.9 & 27.9 \\                                
    5 & (22,22)  & 834 & 3.0 & 22.1 & 18.9 & 18.0 \\                                                                                            
    6 & (26,26)  & 978 & 5.0 & 14.8 & 11.5 & 11.4\\
    7 & (30,30)  &1122 & 10 & 9.5 & 7.2 & 7.7 \\ 
    \multicolumn{7}{c}{\bf cc-pVQZ} \\
    2 & (10,10)  & 790   & 0.0  & 64.5 & 63.1 & 62.2\\      
    3 & (14,14)  & 1070 & 0.0 & 47.0 & 44.4 & 43.2\\                                                                                                  
    4 & (18,18)  & 1350 & 1.0 & 32.8 & 29.2 & 28.3 \\  
   \hline                                                                                                                                                                            
                                                    
   \hline
   \end{tabular*} 
   \label{tab:st}                                                                                                                                                                            
    \begin{tablenotes}
    \footnotesize
    \item${}^{a}$ Number of non-frozen orbitals.
    \item${}^{b}$ The ACI energy importance criteria (m\Eh). See eq \eqref{eq:cond}.
    \end{tablenotes}          
    \end{threeparttable}                                                                                                                                                       
\end{table} 

\begin{table*}[ht!]    
\begin{threeparttable}
    \footnotesize    \caption{Comparison of the best ACI-DSRG-MRPT2 adiabatic singlet-triplet splittings (\kcal) of the acene series with selected literature values}       
    \begin{tabular*}{\textwidth}{l @{\extracolsep{\fill}} rrrrrrrr}   
    \hline  
                     
    \hline            
         &   \multicolumn{6}{c}{$n$-acene} \\
    Method & 2 & 3 & 4 & 5 & 6 & 7 & MUD$^{n}$\\\hline
    UB3LYP$^{\rm{a}}$ & 62.6 & 41.8 & 27.7 & 17.9 & 10.9 & 5.6 & 2.9\\
    CCSD(T)$^{\rm{b}}$ & 65.8 & 48.1 & 33.5 & 25.3 & 17.7 & 13.4 & 3.0 \\
    pp-RPA@U$^{\rm{c}}$ & 66.2 & 45.7 & 32.1 & 22.6 & 15.2 & 9.0 & 1.2 \\
    CAS(8,8)-CISD+Q$^{\rm{d}}$ & 65.5 & 48.4 & 38.5 & 27.7 & 24.2 &16.6 & 5.7 \\                                
    GAS-pDFT (FP-1)$^{\rm{e}}$ & 70.6 & 45.5 & 33.6 & 25.4 & 19.7 & 16.5 & 3.9\\ 
    GAS-pDFT (WFP-3)$^{\rm{e}}$ & 64.7 & 43.1 & 28.8 & 20.5 & 15.0 & 10.0 & 1.4 \\   
    DMRG-pDFT$^{\rm{f}}$ & 67.1 & 46.1 & 31.6 & 22.6 & 16.8 & 14.3 & 1.7\\   
    DMRG-CASPT2$^{\rm{g}}$ & --  & 39.0 & 27.2 & 18.8 &13.3 & -- & 3.2\\               
    DMRG-CASPT2$^{\rm{h}}$ & --  & 39.8 & 29.6 & 19.8 &14.2 & -- & 2.2\\
    This work & 62.2 & 43.2 & 28.3 & 18.0 & 11.4 & 7.7       & 2.5       \\ \\
    ZPVE & $-3.4$ & $-2.3$ & $-1.8$ & $-1.5$ & $-1.3$ & $-1.2$ \\
    Exp.  & $60.9,^{\rm{i}}$ 61.0$^{\rm{j}}$ & 42.6,$^{\rm{i}}$ 43.1$^{\rm{k}}$ & 29.4$^{\rm{i}}$ & $19.8\pm0.7^{\rm{l}}$ & ($12.4\pm1.2)^{\rm{m}}$ & --\\                                                                           
                                                                                                                                                 
   \hline                                                                                                                                                                            
                                                    
   \hline
   \end{tabular*}
   \begin{tablenotes}
   \footnotesize
   \item$^{\rm{a}}$ Ref. \citenum{Hachmann:2007ft}.
   $^{\rm{b}}$ Ref. \citenum{Hajgato:2011jg}.
   $^{\rm{c}}$ Unrestricted geometry. Ref \citenum{Yang15082016}. 
   $^{\rm{d}}$  6-31G basis Ref. \citenum{Horn:2014tu}. 
   $^{\rm{e}}$  tPBE functional and 6-31+G(p,d), Active space partitioning in parentheses defined in Ref. \citenum{Ghosh:2017gl}.
    $^{\rm{f}}$  tPBE functional and 6-31+G(p,d) Ref. \citenum{Sharmadmrgdft}.
   $^{\rm{g}}$  CAM-B3LYP/6-31G* geometry. Ref. \citenum{Kurashige:2014uf}.
   $^{\rm{h}}$   CAS(12,12)-CASPT2-D geometries. Ref. \citenum{Kurashige:2014uf}.
   $^{\rm{i}}$ Ref \citenum{Siebrand:1967dg}.
   $^{\rm{j}}$ Ref \citenum{birks1970}.
   $^{\rm{k}}$ Ref \citenum{Schiedt:1997hy}.
   $^{\rm{l}}$ Ref. \citenum{Burgos:1977ck}.
   $^{\rm{m}}$ Ref. \citenum{Angliker:1982jb}, based on extrapolated correlations of triplet energies to singlet energies and ionization potentials for lower acenes.
   $^{\rm{n}}$ Mean unsigned deviation with respect to vibrationally corrected experimental values.

    \end{tablenotes}                                                                                     
    \label{tab:comp}                                                                                                                                                                   
    \end{threeparttable}		
\end{table*} 

Additionally in Table \ref{tab:comp} we summarize experimental and theoretical predictions of $\Delta E_{\rm{ST}}$ from a variety of methods including DFT, CC, various multireference theories, and ACI-DSRG-MRPT2 best estimates.\cite{Hachmann:2007ft,Hajgato:2011jg,Horn:2014tu,Ghosh:2017gl,Kurashige:2014uf} 
None of the values reported in Table \ref{tab:comp}  include zero-point vibrational energy corrections (ZPVEs), which we show computed with UB3LYP/6-31G($d$).
Aside from UB3LYP, the methods with the least amount of explicitly treated static correlation, 
coupled cluster with singles, doubles, and perturbative triples [CCSD(T)], 
the particle-particle random phase approximation (pp-RPA), 
and CAS(8,8)-CISD+Q, generally overestimate the splittings with errors increasing with acene length. 
When ZPVE corrections are included, ACI-DSRG-MRPT2 consistently underestimate the experimental gaps by 1.7--3.3 \kcal, similar on average 
to the DMRG-CASPT2 results reported by Kurashige and Yanai\cite{Kurashige:2014uf} 
which show larger ($\approx 5$ \kcal) absolute errors for the smaller acenes with respect to experiment.
The DMRG-pDFT predicts singlet-triplet gaps around 2--5 \kcal larger than our results and
are on average closer to the experimental values by 0.8 \kcal compared to ACI-DSRG-MRPT2 despite using a smaller basis set.
Interestingly, DMRG-CASPT2 using the DFT-optimized geometry and GAS-pDFT using an active space partitioning which maximizes the number of determinants included
both agree well with ACI-DSRG-MRPT2 with an average absolute deviation of about 2 \kcal.

As shown by the DMRG-CASPT2 data in Table \ref{tab:comp},
different geometry optimization procedures can cause deviations in 
$\Delta E_{\rm{ST}}$ on the order of 1--3 \kcal, suggesting that a geometry optimization scheme more accurate than DFT is required 
to enable reliable comparisons with experiment. Additionally, most experimental data shown here involve solid or liquid stabilizing
matrices, and in some cases are derived from indirect measurements, both having unpredictable deviations with respect to the zero-temperature, gas-phase results from calculations.
In light of these complications in experimental comparisons, the general agreement of the relaxed ACI-DSRG-MRPT2 and the DMRG-CASPT2 
results of Kurashige and Yanai\cite{Kurashige:2014uf} when both computations use DFT optimized geometries is particularly encouraging
considering that the ACI-DSRG-MRPT2 can be applied to heptacene using a CAS(30,30). Furthermore, the absolute errors of the
ACI-DSRG-MRPT2 with respect to experiment do not systematically deviate as a function of acene length, with all errors within a 1.6 \kcal window. This observed
consistency would predict an experimental singlet-triplet gap for heptacene to be roughly 2.5 \kcal above our reported value.

\subsection{Emergent Radical Character}
One of the greatest benefits of our approach is that we can analyze the importance of relaxation effects on electronic properties.
We emphasize that the relaxed wave functions and densities we present span the $\pi/\pi^{*}$ active spaces used in building the
references, and they are optimized in \emph{response} to dynamical effects rather than being built in the full orbital basis of the g
DSRG-MRPT2.
To investigate the emergent radical character, and the importance of relaxation effects in accurately describing it, we compute
the effective number of unpaired electrons as defined by Takatsuka et al.,\cite{Takatsuka:1978gk,HeadGordon:2003kq}
\begin{equation}
\text{ \# of unpaired electrons } =\sum_{i} n_{i}(2-n_{i})
\end{equation}
for each natural orbital occupation number ($n_{i}$)
computed from relaxed and unrelaxed active space densities.
In our wave function analysis, we use a constant $\sigma$ per number of electrons in order to ensure that each 
acene is computed to the same relative accuracy, so that our interpretation of trends is unaffected by any
potential differences in reference wave function quality.

\begin{figure}[ht!]
\centering
\includegraphics[width=3.325in]{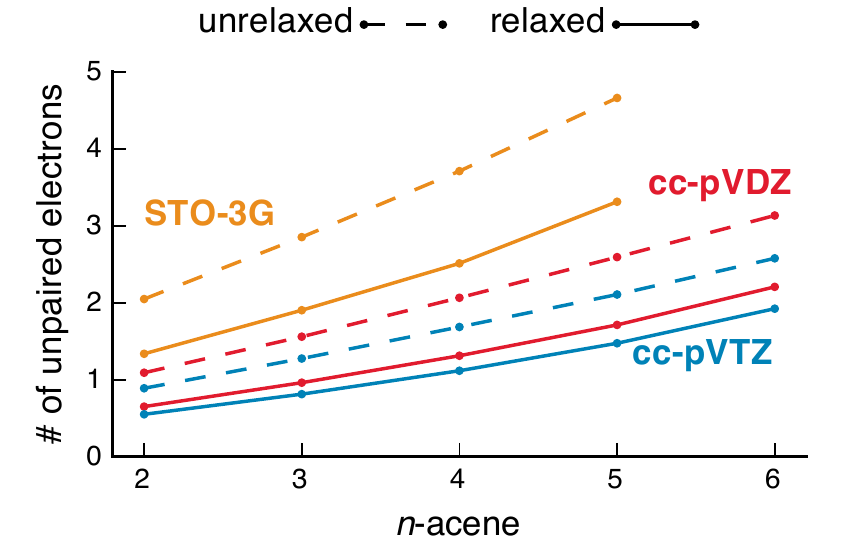}
\caption{Number of unpaired electrons for the ground state singlets of the oligoacenes computed from unrelaxed (dashed line) and relaxed (solid line) ACI-DSRG-MRPT2 wave functions. Sigma values are chosen to produce a constant 190 $\mu E_{\rm{h}}$ error per electron.}
\label{fig:nos}
\end{figure}

Figure \ref{fig:nos} shows the effective number of unpaired electrons for the acene series computed from unrelaxed and relaxed 
references of ground state singlets, using STO-3G, cc-pVDZ, and cc-pVTZ basis sets.
Improving the dynamical correlation treatment by both increasing
the basis set and relaxing the reference reduces the observed radical character dramatically. 
When used in small basis sets, this metric yields only qualitative information about the relative radical character among acenes
and is unable to provide any definitive insight into when the degree of radical character is significant. 
In computing this metric more accurately, we see quantitative evidence for the emergent diradical character in hexacene, though
we are cautious to map this metric directly to a chemical observable.
While increasing the correlation treatment and basis set quality is likely to further decrease this metric, 
we already see relatively close agreement between results using cc-pVTZ and cc-pVDZ basis sets.
Our observation of a slower emergence of radical
character is consistent with the previously reported notion\cite{Lee:2017bh,Battaglia:2017bb,Dupuy:2018gp} that 
small basis sets and only an active-space treatment of electron correlation can lead to overestimating the radical character and misinterpreting the nature of the ground state. Ultimately, ACI-DSRG-MRPT2 results do show weak emergent radical character with acene length.

\begin{figure}[ht!]
\centering
\includegraphics[width=3.25in]{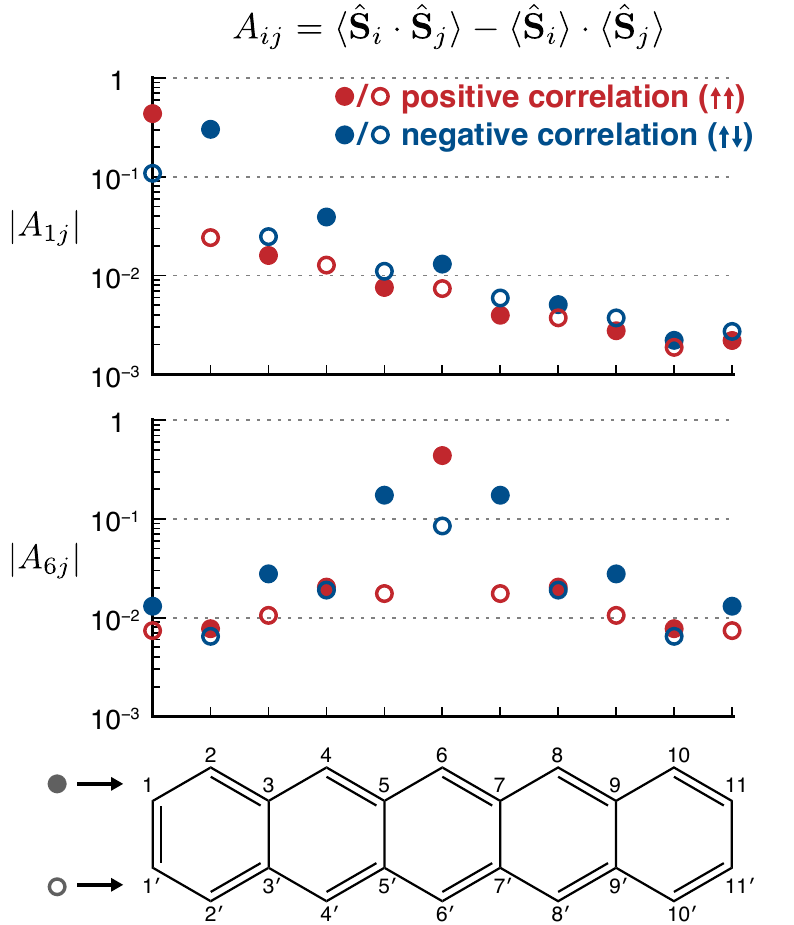}
\caption{Log-plot of the spin-spin correlation function using carbons 1 and 6 as references,
 computed for the ground state singlet of pentacene with a relaxed ACI-DSRG-MRPT2 wave function.
}
\label{fig:spin_plot}
\end{figure}

\subsection{Analysis of Spin-Spin Correlation}
As a final analysis, we characterize spatial correlations of spin by computing the spin-spin correlation function ($A_{ij}$) between two atomic sites $i$ and $j$, defined as
$A_{ij} = \langle \hat{\mathbf{S}}_{i} \cdot \hat{\mathbf{S}}_{j}\rangle - \langle \hat{\mathbf{S}}_{i}\rangle\cdot\langle\hat{\mathbf{S}}_{j}\rangle$,
where each site is defined as an atom-centered, Pipek--Mezey localized\cite{Pipek:1989ci} carbon $2p_{z}$-like molecular orbital
, 
and $\hat{\mathbf{S}}_{i}$ is the total spin operator for site $i$. 
This correlation function depends on the 1- and 2- RDMs of the ACI wave functions, enabling us to understand the effect of reference relaxation on the spin correlations, though it will not include spin correlations outside the $\pi/\pi^{*}$ manifold.
As shown in Figure \ref{fig:spin_plot}, spin-spin correlation in pentacene is large at small distances but quickly decays with a pattern characteristic of short-range antiferromagnetic order.
Interestingly, opposite-spin correlations are larger than same-spin correlations along the same edge, but have similar magnitudes along for
the opposite edge. This result indicates that any biradical character in the acenes is localized along the long axes of the molecule and stabilized
with antiferromagnetic ordering, confirming previous hypotheses.\cite{Dupuy:2018gp}

\begin{figure}[ht!]
\centering
\includegraphics[width=3.25in]{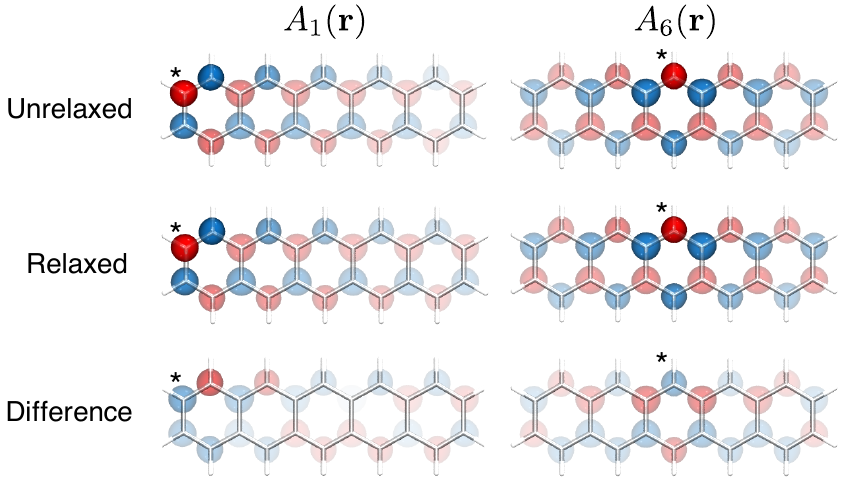}
\caption{Spin correlation densities plotted from unrelaxed and relaxed references, in addition to the difference of relaxed and unrelaxed results. We show two reference sites, marked with ``*'', corresponding to central and terminal carbons consistent with the labeling in Figure \ref{fig:spin_plot}, with all reference sites plotted in the SI.}
\label{fig:spin_density}
\end{figure}

In addition, we also compute the spin-spin correlation density with respect to site $i$ [$A_{i}(\mathbf{r})$], defined as,
\begin{equation}
A_{i}(\mathbf{r}) =
\langle \hat{\mathbf{S}}(\mathbf{r}) \cdot \hat{\mathbf{S}}_{j}\rangle - \langle \hat{\mathbf{S}}(\mathbf{r})\rangle\cdot\langle \hat{\mathbf{S}}_{j}\rangle \approx \sum_{j} |\phi_{j}(\mathbf{r})|^{2}A_{ij}
\end{equation}
where $\hat{\mathbf{S}}(\mathbf{r})$ is the total spin operator in real space and the last approximate equality assumes that overlap terms can be neglected due to orbital localization.
In Figure~\ref{fig:spin_density}, we use this metric to illustrate the spatial distribution and the effect of reference relaxation on spin-spin correlation in pentacene.
Dynamical correlation generally increases same and opposite spin correlations for long 
range interactions, while decreasing the short-range opposite correlations indicative of bonding. Thus, the relaxation
effects enhance long-range effects while reducing short-range ones.
Note that our plots of the spin correlation density do not explicitly reflect open shell character, but they do show short-range antiferromagnetic order. 

\section{Conclusions}
In this work, we have introduced the ACI-DSRG-MRPT2 method for studying large-scale strong correlation. Our approach uses
the efficient, tunable ACI to recover static correlation within large active spaces and the intruder-free 
DSRG-MRPT2 to recover dynamical correlation, which we can use to recompute active space wave functions with consideration of dynamical correlation.
We have shown that this strategy can be applied to large active spaces, up to CAS(30,30), and for systems using up to 1350 basis functions on a single node.
We find that this procedure provides accurate energies and high-quality wave functions
suitable for quantitative analysis. 
Our application to the oligoacenes has demonstrated that relaxation effects can significantly influence interpretation of chemical
properties, with notable reduction of radical character and a shifting of spin correlations to longer distance.

In addition, the ACI-DSRG-MRPT2 is flexible. With reference wave functions formed
from determinants, we can easily apply our approach to excited states using state-specific or multistate
DSRG approaches.\cite{Li:2017ff}
Furthermore, we are not limited to second order perturbation theory, as the higher-order non-perturbative variants
of the DSRG still only require at most the 3-RDM.
With further development of ACI, we can apply ACI-DSRG theories to ground and excited states of even larger,
more complex molecules.

\section{Acknowledgments}
This work was supported by the U.S. Department of Energy under Award No. DE-SC0016004 and a Research Fellowship of the Alfred P. Sloan Foundation.


\providecommand{\latin}[1]{#1}
\providecommand*\mcitethebibliography{\thebibliography}
\csname @ifundefined\endcsname{endmcitethebibliography}
  {\let\endmcitethebibliography\endthebibliography}{}

\end{document}